\documentclass[epj]{svjour}
\usepackage{graphics}
\newcommand{\lapprox}{%
\mathrel{%
\setbox0=\hbox{$<$}
\raise0.6ex\copy0\kern-\wd0
\lower0.65ex\hbox{$\sim$}
}}
\newcommand{\gapprox}{%
\mathrel{%
\setbox0=\hbox{$>$}
\raise0.6ex\copy0\kern-\wd0
\lower0.65ex\hbox{$\sim$}
}}

\newcommand{\beq}{\begin{equation}} 
\newcommand{\eeq}{\end{equation}} 
\newcommand{\beqa}{\begin{eqnarray}} 
\newcommand{\eeqa}{\end{eqnarray}} 
\newcommand{\beqan}{\begin{eqnarray*}} 
\newcommand{\eeqan}{\end{eqnarray*}} 
\newcommand{\ba}{\begin{array}} 
\newcommand{\ea}{\end{array}} 
\newcommand{\no}{\nonumber}

\newcommand{\wt}{\widetilde}

\newcommand{\D}{{\cal D}} 
 
\newcommand{\F}{{\cal F}}

\newcommand{\cL}{{\cal L}}

\newcommand{\Q}{{\cal Q}}

\newcommand{\nn}{\nonumber \\}

\newcommand{\bea}{\begin{eqnarray}} 
\newcommand{\eea}{\end{eqnarray}}

\begin{document}
\title{The pionic beta decay in chiral perturbation theory
\thanks{Work supported in part by IHP-RTN, EC-Contract  
No. HPRN-CT-2002-00311 (EURIDICE). }
}
\author{V. Cirigliano$^1$, M. Knecht$^2$, H. Neufeld$^3$, 
H. Pichl$^4$ 
}                     
%
%
\institute{
$^1$ Departament de F\'{\i}sica Te\`{o}rica, IFIC, Universitat de 
Val\`{e}ncia -- CSIC, Apt. Correus 22085, E-46071 Val\`{e}ncia, Spain \\
$^2$ Centre de Physique Th\'eorique, CNRS 
Luminy, Case 907, F-13288 Marseille - Cedex 9, France \\
$^3$ Institut f\"ur Theoretische Physik, Universit\"at 
Wien, Boltzmanngasse 5, A-1090 Wien, Austria \\
$^4$ Paul Scherrer Institut, CH-5232 Villigen PSI, Switzerland
}
\date{ Received: date / Revised version: date     }
%
\abstract{Within the framework of chiral perturbation theory with 
virtual photons and leptons, we present an updated analysis of the pionic 
beta decay including all electromagnetic contributions of order $e^2 p^2$. 
We discuss the extraction of the Cabibbo--Kobayashi--Maskawa matrix element 
$|V_{ud}|$ from  experimental data. The method employed here is consistent 
with the analogous treatment of the $K_{\ell 3}$ decays and the 
determination of $|V_{us}|$.
} 
\maketitle
%
%


\section{Introduction}
\label{sec: Introduction}
\renewcommand{\theequation}{\arabic{section}.\arabic{equation}}
\setcounter{equation}{0}


The determination of the quark mixing matrix $V_{\rm CKM}$ is
presently a major issue in elementary particle physics. The existence
of a direct CP violating phase in the kaon sector has now been clearly
established experimentally \cite{PDG}. 
Furthermore, the 
failure of the CKM matrix elements to satisfy the constraints
expressing the unitarity of $V_{\rm CKM}$ would with certainty
establish the existence of new degrees of freedom beyond those
present in the three generation standard model.  
Since $\vert V_{ub}\vert \lapprox 5\times 10^{-3}$ (the latest value
obtained by CLEO from inclusive semileptonic $B$ decays reads
$\vert V_{ub}\vert = (4.08\pm0.63)\times 10^{-3}$ \cite{CLEO}) and 
given the present accuracies on $\vert V_{ud}\vert$ and $\vert V_{us}\vert$,
at the 0.1\% and 1\% level, respectively, the most
stringent test of the unitarity of the CKM matrix comes from the light quark 
sector, 
where $\vert V_{ud}\vert^2 \,+\, \vert V_{us}\vert^2$ should show
compatibility with unity with an accuracy better than 0.3\%.
The semileptonic $K_{e3}$ decay modes are presently considered to provide 
the best determination of $\vert V_{us}\vert$ with the current value 
$\vert V_{us}(K_{e3})\vert=0.2196\pm 0.0026$ \cite{PDG}, 
while the most accurate measurement of $\vert V_{ud}\vert$ relies 
on the $\F{\mbox t}$ 
values obtained from the super-allowed Fermi transitions of several 
$0^{+}$ nuclei, $\vert V_{ud}(\F{\mbox t})\vert =0.9740\pm 0.0005$ 
\cite{WEIN98}. This gives 
$\vert V_{ud}(\F{\mbox t})\vert^2 \,+\, \vert V_{us}(K_{e3})\vert^2 
= 0.9969(17)$, i.e.
a deviation from unity by 2$\sigma$. It is of course too early
to draw definite conclusions from this result. On the one hand, the value
of $\vert V_{ud}\vert$ obtained from the nuclear beta decays hinges on the
control of the nuclear structure aspects involved in the evaluation of 
the radiative corrections to the transition matrix elements. 
On the other hand, the extraction 
of $\vert V_{us}\vert$ from the 
$K_{e3}$ decay modes relies on a one-loop chiral perturbation calculation 
\cite{lr84,CKNRT} 
of the relevant form factor, supplied with a model dependent 
estimate of the higher order contributions. The Particle Data Group 
compilation therefore recommends to double the uncertainty in the value 
of $\vert V_{ud}\vert$ inferred from the nuclear Fermi transitions,  
$\vert V_{ud}(\F{\mbox t})\vert =0.9740\pm 0.0010$. 
In the case of the $K_{e3}$ decay mode,
other quark model estimates of higher order corrections lead to a
wider range of values than those considered in \cite{lr84}.
A similar conclusion, based on the analysis of ${\cal O}(m_s^2)$ tree-level
corrections in both $F_K/F_\pi$ and $K_{e3}$ form factors,
has also been reached in \cite{fks00}\footnote{The two-loop
expressions of the $K_{\ell 3}$ form factors have been worked out
in \cite{ps01}, but no evaluation of the relevant 
${\cal O}(p^6)$ counterterms was given.}. 
Determinations of $\vert V_{ud}\vert$ from other sources than the nuclear 
Fermi transitions have therefore been considered. The neutron
beta decay, where radiative corrections can be evaluated in a 
more reliable way,  suffers from the drawback that in this case 
both the vector current and the axial-vector
current contribute. It is thus not enough  to measure the lifetime only, 
but the asymmetry in the electron emission angle with respect to 
the neutron polarization is also required. A recent measurement of
this asymmetry \cite{nbeta02} leads to a value 
$\vert V_{ud}(n\beta)\vert =0.9713\pm 0.0013$, even smaller than the one 
coming from the superallowed nuclear $0^+ \to 0^+$ transitions, and 
corresponds to 
$\vert V_{ud}(n\beta)\vert^2 \,+\, \vert V_{us}(K_{e3})\vert^2 
= 0.9916(28)$, i.e.
a 3$\sigma$ deviation from unitarity. 

Another interesting possibility, 
which shares the advantages of both Fermi transitions (pure vector transition, 
no axial-vector contribution) and neutron beta decay (no nuclear structure
dependent radiative corrections) is provided by the beta decay of the 
charged pion. The difficulty here lies in the extremely small branching 
ratio, $\sim 10^{-8}$. Nevertheless, such a measurement is presently
being performed at PSI by the PIBETA collaboration \cite{PSI}, with the 
aim of 
measuring the branching ratio with 0.5\% accuracy. At this level of 
precision, radiative corrections have to be taken into account, and 
the present paper addresses this problem. We shall use the effective 
theory formalism for processes involving light pseudoscalar mesons, 
photons and leptons introduced in \cite{lept} which is particularly
well suited for the pionic beta decay.
After a brief review of the main kinematic features of the process 
(Section 2), we describe the modifications to the structure of the 
decay amplitude induced by the radiative corrections in Section 3.
In particular, we obtain the corrections of order ${\cal O}(e^2p^2)$
to the form factor. Section 4 discusses the radiative beta decay 
rate, which has to be included in order to cancel the infrared 
divergences which appear in the radiative corrections to the 
beta decay amplitude without real photon emission. Numerical 
estimates, in particular of the theoretical uncertainty in the 
determination of $\vert V_{ud}(\pi \beta)\vert$, are presented in Section 
5.
Conclusions are given in Section 6, and two Appendices contain some
technical material related to the calculation of the loop contributions.


\section{Kinematics}
\label{sec: Kinematics} 
\renewcommand{\theequation}{\arabic{section}.\arabic{equation}}
\setcounter{equation}{0}

In the absence of radiative corrections,  the invariant amplitude of the 
decay
\beq
\pi^+ (p_+) \to \pi^0 (p_0) \, e^+ (p_e) \, \nu_e (p_\nu) 
\eeq
reads
\beq
{\cal M} = 
G_{\rm F} V_{ud}^{*} \, l^{\mu}  
\bigg[ f_{+}^{(0)} (t) \, (p_+ + p_0)_{\mu} 
+  f_{-}^{(0)} (t) \, (p_+ - p_0)_{\mu} \bigg]  ,
\label{basic1}
\eeq
where 
\beq
l^\mu = \bar{u} (p_\nu) \, \gamma^\mu \, (1 - \gamma_5) \, v (p_e) .
\eeq
The expression in parentheses corresponds to the matrixelement $\langle 
\pi^0(p_0) | \bar{u} \gamma_\mu d | \pi^+(p_+) \rangle / \sqrt{2}$.
The hadronic form factors depend on the single variable 
$t = (p_+ - p_0)^2$. 

The spin-averaged decay distribution $\rho(y,z)$ is a function of the 
two variables 
\beq \label{defyz}
z = \frac{ 2 p_+ \cdot p_0 }{M_{\pi^+}^2} = 
\frac{2 E_{\pi^0}}{M_{\pi^+}}  ,  \quad y = \frac{ 2 p_+ \cdot p_e 
}{M_{\pi^+}^2} =
\frac{2 E_{e^+}}{M_{\pi^+}}  , \eeq
where $E_{\pi^0}$ ($E_{e^+}$) is the $\pi^0$ (positron) energy in 
the rest frame of the charged pion.
Alternatively, one may also use two of the Lorentz 
invariants 
\beqa
t &=& (p_+ - p_0)^2 = M_{\pi^+}^2 (1 + r_\pi -z), \nn
u &=& (p_+ - p_e)^2 = M_{\pi^+}^2 (1 + r_e - y), \nn 
s &=& (p_0 + p_e)^2 = M_{\pi^+}^2 (y + z - 1),  
\eeqa
where
\beq
r_e = \frac{m_e^2}{M_{\pi^+}^2}  , \qquad r_\pi = 
\frac{M_{\pi^0}^2}{M_{\pi^+}^2}. 
\eeq
Then the Dalitz plot density (without radiative corrections) reads 
\beqa
\rho^{(0)} (y,z) &=&  {\cal N}
\Big[ A_1^{(0)}  |f_{+}^{(0)} (t)|^2  
+ A_2^{(0)}   f_{+}^{(0)} (t)   f_{-}^{(0)} (t) \nn 
&& \quad {}+ A_3^{(0)}   |f_{-}^{(0)} (t)|^2 \Big]  ,  
\label{basic2}
\eeqa
with  
\beq
{\cal N}  = \frac{G_{\rm F}^2 \, |V_{ud}|^2 M_{\pi^+}^5}{64 \pi^3}  , 
\eeq
\beq
\quad \Gamma(\pi^+ \to \pi^0 e^+ \nu_e) = \int\limits_{\cal D}  dy \, dz \ 
\rho^{(0)} (y,z)  .  
\eeq
The kinematical densities are given by 
\beqa
A_1^{(0)} (y,z) &=& 4 (z + y - 1) (1 - y) 
+ r_e (4 y + 3 z - 3) \no \\
&&{}- 4 r_\pi + r_e (r_\pi - r_e)    , \nonumber \\
A_2^{(0)} (y,z) & = & 2 r_e (3 - 2 y - z + r_e - r_\pi)  , \nn
A_3^{(0)} (y,z) & = &  r_e ( 1 + r_\pi - z - r_e)   . 
\eeqa

The physical domain ${\cal D}$ is defined by 
\beqa \label{domain1}
2 \sqrt{r_e} & \leq y \leq & 1 + r_e - r_\pi  , \nonumber \\ 
a (y) - b (y) & \leq z \leq & a (y) + b (y)  ,  
\eeqa
where  
\beqa \label{domain2}
a (y) & = & \frac{(2 - y) \, (1 + r_e + r_\pi -y)}{2 ( 1 + r_e - y)}
 , \nonumber \\ 
b (y) & = & \frac{\sqrt{y^2 - 4 r_e} \, 
(1 + r_e - r_\pi -y)}{2 ( 1 + r_e - y)} ,   
\eeqa
or, equivalently,
\beqa \label{domain3}
2 \sqrt{r_\pi} & \leq z \leq & 1 + r_\pi - r_e  , \nonumber \\ 
c (z) - d (z) & \leq y \leq & c (z) + d (z)  ,  
\eeqa
where  
\beqa \label{domain4}
c (z) & = & \frac{(2 - z) \, (1 + r_\pi + r_e -z)}{2 ( 1 + r_\pi - z)}
 , \nonumber \\ 
d (z) & = & \frac{\sqrt{z^2 - 4 r_\pi} \, 
(1 + r_\pi - r_e -z)}{2 ( 1 + r_\pi - z)}  . 
\eeqa

\section{Pionic beta decay and electromagnetism}
\label{sec: elm} 
\renewcommand{\theequation}{\arabic{section}.\arabic{equation}}
\setcounter{equation}{0}

The presence of the electromagnetic interaction  does not 
change the structure of the invariant amplitude (\ref{basic1}) in terms of 
the form factors, but changes the form factors themselves \cite{CKNRT}:
\beq
{\cal M} [f_+^{(0)}(t),f_-^{(0)}(t)] \to
{\cal M} [F_+(t,u),F_-(t,u)]. 
\label{shift}
\eeq
The full form factors $F_{\pm}(t,u)$ contain the effects of virtual 
photon exchange and the contributions of the appropriate 
electromagnetic counterterms. These quantities depend also on a second
kinematical variable as they cannot be interpreted
anymore as matrix elements of a quark current between hadronic
states. 
%
%

\begin{figure}[h]
\hspace*{2.0cm}
\resizebox{0.25\textwidth}{!}{
  \includegraphics{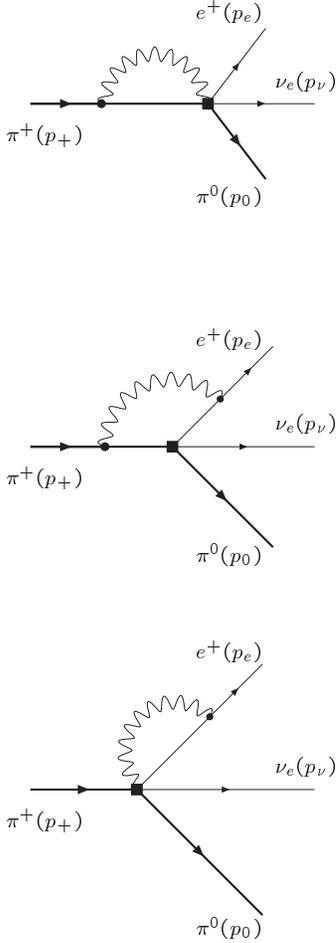}}
\caption{Photon loop diagrams (wave function renormalization
diagrams are not shown). The black dots denote (scalar) QED 
vertices, the box denotes the vertex proportional to $G_F$. }
\label{pb-diagrams}    
\end{figure}

Photon loop diagrams contributing to the weak vertex function are shown in 
Figure \ref{pb-diagrams}. In particular, it is the second diagram which 
introduces the dependence on the second kinematical variable $u$. 
%
%

The following observation considerably simplifies the further analysis of 
the pionic beta decay: the form factor $F_-(t,u)$ is proportional
to the isospin suppressed mass difference $M_{\pi^+}^2 - M_{\pi^0}^2 = 2 
e^2 Z F_0^2$. In addition, the kinematical densities $A_2$ and $A_3$ 
multiplying $F_-$ in the formula for the Dalitz plot density 
(\ref{basic2}) are proportional to the small quantity 
$r_e \simeq 10^{-5}$. Therefore, these contributions can safely be 
neglected and we may restrict ourselves to the discussion of $F_+$.

The form factor $F_+(t,u)$ contains infrared singularities due to 
low-momentum virtual photons. They can be regularized  by introducing a 
small photon mass 
$M_\gamma$.  The dependence on an infrared cutoff reflects the fact that 
$F_+(t,u)$ cannot be interpreted as an observable quantity but has to be 
combined with appropriate contributions from real photon emission to 
arrive at an infrared-finite result. 

It is convenient to decompose $F_+(t,u)$ into a structure-dependent 
effective form factor $f_+(t)$, and a remaining part containing in 
particular the universal long distance corrections. Confining ourselves to 
electromagnetic contributions of order $\alpha$, the full form factor is 
given by
\beq 
F_+ (t,u) = \left[ 1 + \frac{\alpha}{4 \pi} 
\Gamma (u,m_e^2,M_{\pi^+}^2;M_\gamma) \right] \ f_+ (t)   . 
\label{factor1}
\eeq 
Expressed in terms of the functions $\Gamma_c$, $\Gamma_1$, $\Gamma_2$ 
defined in \cite{CKNRT}, $\Gamma$ can be written as
\beqa
\Gamma(u,m_e^2,M_{\pi^+}^2;M_\gamma) &=& 
\Gamma_c(u,m_e^2,M_{\pi^+}^2;M_\gamma) \nn *
&& {} + \Gamma_1(u,m_e^2,M_{\pi^+}^2) \nn *
&& {} + \Gamma_2(u,m_e^2,M_{\pi^+}^2) .
\eeqa
The explicit expressions of the functions $\Gamma_c$, $\Gamma_1$, $\Gamma_2$ 
can be found in Appendix \ref{appA}. 
$\Gamma_c(u,m_e^2,M_{\pi^+}^2;M_\gamma)$ corresponds to the long distance 
component of the loop amplitudes which generates infrared and Coulomb 
singularities. (In our case, the Coulomb singularity lies outside of the 
physical region.) The term $\Gamma_1(u,m_e^2,M_{\pi^+}^2) + 
\Gamma_2(u,m_e^2,M_{\pi^+}^2)$ represents the remaining nonlocal photon 
loop contribution .

Note that the decomposition in (\ref{factor1}) 
is analogous to the one chosen in \cite{CKNRT} for the analysis of the 
$K_{e3}^+$ decay. With this choice, the effective form factor depends only 
on the single variable $t$. It is convenient \cite{CKNRT} to write it as 
the sum of two terms,
\beq \label{splitting}
f_+(t) = \wt{f}_+(t) + \left. f_+ \right|_{e^2 p^2}.
\eeq
The first one contains the pure QCD contribution (in principle to any 
order in the chiral expansion) plus the electromagnetic contributions 
originating from the non-derivative Lagrangian
\beq
\cL_{e^2 p^0} = e^2 F_0^4 Z \langle \Q_L^{\rm em} \Q_R^{\rm em} \rangle .
\eeq
The explicit form of this part is given by the formula
\beq \label{fplus-wiggle}
\wt{f}_{+} (t) = 1  +  2 H_{\pi^+ \pi^0} (t) +  H_{K^+ K^0} (t) + 
\dots ,
\eeq
where the ellipses indicate contributions of higher order in the chiral 
expansion.
The meson loop function $H_{PQ}(t)$ \cite{gl852,GL85} is displayed in 
Appendix \ref{appB}. 

The second term in (\ref{splitting}) represents the local effects of 
virtual photon exchange of order $e^2 p^2$, 
\begin{eqnarray}
\left. f_{+} \right|_{e^2 p^2} & = & 
4\pi\alpha \Big[
2K_{12}^r(\mu) - \frac{2}{3} X_1 - \frac{1}{2} \wt{X}_6^r(\mu) 
\nn
&-& \frac{1}{32\pi^2} \Big(
3 + \log\frac{m_e^2}{M_{\pi^+}^2} + 3\log\frac{M_{\pi^+}^2}{\mu^2}
\Big)\Big] . 
\label{fplus-em-loc}
\end{eqnarray}
The symbol $K_{12}^r(\mu)$ denotes the renormalized 
(scale dependent) part of the coupling constant $K_{12}$ introduced by 
Urech \cite{urech} in the effective Lagrangian $\cL_{e^2 p^2}$ describing 
the interaction of dynamical photons with hadronic degrees of freedom 
\cite{nr95,nr96}. The coupling constants $X_1, X_6$ enter the game once 
also virtual leptons are taken into account \cite{lept}. The coupling 
constant $\wt{X}_6^r(\mu)$ is obtained from  $X_6^r(\mu)$ after 
the subtraction of the short distance contribution \cite{CKNRT},
\beq
X_6^r(\mu) = X_6^{\rm SD} + \wt{X}_6^r(\mu),
\eeq
where
\beq
e^2 X_6^{\rm SD} = -\frac{e^2}{4 \pi^2} \log \frac{M_Z^2}{M_\rho^2}
= 1 - S_{\rm EW}(M_\rho, M_Z),
\eeq
which defines \cite{MS93} also the short distance enhancement factor 
$S_{\rm EW}(M_\rho, M_Z)$ to leading order.

In order to arrive at an infrared-finite (observable) result, also the 
emission of a real photon has to be taken into account.
The radiative amplitude ${\cal M}^\gamma$ can be expanded in powers of the 
photon energy $E_\gamma$, 
\beq 
{\cal M}^\gamma = {\cal M}_{(-1)}^{\gamma}
+ {\cal M}_{(0)}^{\gamma} + \dots  , 
\label{low1}
\eeq 
where 
\beq {\cal
M}_{(n)}^{\gamma} \sim (E_\gamma)^n   .  
\eeq 
Gauge invariance relates ${\cal M}_{(-1)}^{\gamma}$ 
and ${\cal M}_{(0)}^{\gamma}$ to the 
non-radiative amplitude ${\cal M}$, and thus to the full form factor 
$F_+(t,u)$. 
Upon taking the square modulus and summing over spins, the radiative
amplitude generates a correction $\rho^\gamma (y,z)$ to the Dalitz
plot density of (\ref{basic2}). The observable distribution is 
now the sum 
\beq 
\rho (y,z) =  \rho^{(0)} (y,z) \, + \rho^\gamma  (y,z)   .  
\eeq
Both terms on the right hand side of this equation depend on the full
form factor $F_+$ and contain infrared divergences (from virtual or real
soft photons). Upon combining them, the observable density can be
written in terms of a new kinematical density $A_1$ \cite{CKNRT}, and the 
effective form factors $f_+ (t)$ defined in (\ref{factor1}), 
\beq \label{fulldensity}
\rho (y,z) = {\cal N} \,  S_{\rm EW}(M_\rho,M_Z) \,    
A_1  \, |f_{+} (t)|^2 ,  
\eeq
where we have pulled out the short-distance enhancement factor. 
The kinematical density $A_1$ is given by \cite{CKNRT}
\beq \label{deltas}
A_1 (y,z) = A_1^{(0)} (y,z) \,  \left[ 1 + \Delta^{\rm IR} (y,z) \right] 
\,  + 
 \, \Delta_{1}^{\rm IB} (y,z)   . 
\eeq
The function $\Delta^{\rm IR} (y,z)$ arises by combining the contributions 
from $|{\cal
M}_{(-1)}^{\gamma}|^2$ and $\Gamma (u,m_e^2,M_{\pi^+}^2; M_\gamma)$. 
Although the individual contributions contain infrared divergences, the 
sum is  finite.  
The factor $\Delta^{\rm IB}_1 (y,z)$ originates from averaging the 
remaining terms of $|{\cal M}^{\gamma}|^2$ [see (\ref{low1})]  
and are infrared-finite. Note that both $\Delta^{\rm IR} (y,z)$ and 
$\Delta^{\rm IB}_1 (y,z)$ are sensitive to the treatment 
of real photon emission in the experiment. 
Details on these corrections are given in Sect.~\ref{sec: corr}.
 
Let us finally note that, in principle, the radiative amplitude
generates new terms in the density, proportional to derivatives of
form factors.  These terms would only arise at order $e^2 p^4$ and
higher in chiral perturbation theory, and therefore we have suppressed 
them in
(\ref{fulldensity}).

\section{Real photon radiation}
\label{sec: corr}
\renewcommand{\theequation}{\arabic{section}.\arabic{equation}}
\setcounter{equation}{0}

We present here in detail a possible treatment of the contribution of the 
real photon emission process
\beq
\pi^+(p_+) \to \pi^0 (p_0) \,  e^+ (p_e) \, 
\nu_e (p_\nu) \, \gamma (p_\gamma)  ,
\eeq
in complete analogy with the procedure proposed in \cite{gin67} 
and \cite{CKNRT} for the analysis of the $K_{e3}^+$ decay. To this end we 
define the kinematical variable
\beq \label{x}
x = (p_\nu + p_\gamma)^2 = (p_+ - p_0 - p_e)^2  . 
\eeq
For the analysis of the experimental data, we suggest to accept all pion 
and charged lepton energies
within the whole $\pi_{e3}$ Dalitz plot $\D$ given by (\ref{domain1}) and 
(\ref{domain3}), respectively, and all kinematically 
allowed values of the Lorentz invariant $x$ defined in (\ref{x}). 
(The variable $x$ determines the angle between the momentum of the neutral 
pion and the positron momentum for given energies $E_{\pi^0}$, $E_{e^+}$.)
This translates into the distribution 
\beqa \label{rhogamma}
\lefteqn{\rho^\gamma (y,z) =  \frac{M_{\pi^+}}{2^{12}\pi^5} 
\int\limits_{M^2_\gamma}^{x_{\rm max}} 
dx} \nonumber \\*
& \times & \! \frac{1}{2 \pi} \! \int \! \frac{d^3 p_\nu}{p_\nu^0} 
\frac{d^3 p_\gamma}{p_\gamma^0} 
\delta^{(4)}(p_+-p_0-p_e-p_\nu-p_\gamma) \! 
\sum_{\rm pol}  |{\cal M}^{\gamma}|^2   , \nonumber \\*
&&
\eeqa 
with
\beqa \label{xmax} 
x_{\rm max} &=& M_{\pi^+}^2 
\bigg\{1 + r_\pi + r_e - y - z  \nn
&&{}+ \frac{1}{2} \Big[y z  +  \sqrt{(y^2 - 4 r_e)(z^2 - 4 r_\pi)} 
\Big] \bigg\} . 
\eeqa
In (\ref{rhogamma}) we have extended the integration over the whole range 
of the invariant mass of the unobserved $\nu_\ell \, \gamma$ system.
The integrals occurring in (\ref{rhogamma}) have the general form 
\cite{gin67}
\beqa \label{Imn}
\lefteqn{I_{m,n}(p_1,p_2;P,M_{\gamma}) :=} \nonumber \\* 
&& \frac{1}{2 \pi} \int 
\frac{d^3 q}{q^0} \frac{d^3 k}{k^0} 
\frac{\delta^{(4)}(P-q-k)} 
{(p_1 \cdot k 
+M_{\gamma}^2/2)^m (p_2 \cdot k + M_{\gamma}^2/2)^n}  . \nonumber \\*
&&
\eeqa
The results for these integrals in the limit $M_\gamma = 0$ can be found 
in the Appendix of \cite{gin67}. 
Using the definition (\ref{Imn}), the  radiative decay 
distribution (\ref{rhogamma}) can be written as 
\cite{gin67}
\beqa \label{decomp}
\lefteqn{\rho^\gamma (y,z) =} \nonumber \\* 
&& \frac{\alpha}{\pi} \Bigg[ \rho^{(0)} (y,z) I_0(y,z;M_\gamma) 
+\frac{G_{\rm F}^2 |V_{ud}|^2 |f_{+}|^2 M_{\pi^+}}{32 \pi^3} \nn
&& \times \! \! \int\limits_0^{x_{\rm max}} \! \! dx 
\sum_{m,n} c_{m,n} \, I_{m,n}(p_e,p_+;p_+-p_0-p_e,0) \Bigg]  ,
\nonumber \\*
&&
\eeqa
where the infrared divergences are now confined to\footnote{The right-hand 
side of the corresponding expression (6.7) in \cite{CKNRT} should be 
multiplied by $1/4$.} 
\beqa \label{I0}
\lefteqn{I_0(y,z;M_\gamma) =}  \nn
&& \frac{1}{4} \int\limits_{M^2_{\gamma}}^{x_{\rm max}} dx 
\Big[- 2 \, p_+ \cdot p_e \, 
I_{1,1}(p_e,-p_+;p_+-p_0-p_e;M_\gamma) 
\nn[-14pt]
&& \qquad \quad {} - M_{\pi^+}^2 \, 
I_{0,2}(p_e,-p_+;p_+-p_0-p_e;M_\gamma)
\nn
&& \qquad \quad {}- m_e^2 \,
I_{2,0}(p_e,-p_+;p_+-p_0-p_e;M_\gamma) \Big] . 
\eeqa
The explicit form of the function $I_0$ can be found in 
Eq. (27) of \cite{gin67}. (Of course, the appropriate substitutions $K \to 
\pi^+$ and $\pi \to \pi^0$ have to be performed.) 
The coefficients $c_{m,n}$ were given in Eq. (19) of \cite{gin67}. Note 
however the misprint for the values of $c_{-1,0}$ and $c_{1,-2}$ (see 
Erratum of \cite{gin67}). As we are neglecting the contribution of the 
form factor $f_{-}$, it is sufficient to consider these coefficients for 
$\xi = 0$.

The function $\Delta^{\rm IR}$ introduced in (\ref{deltas}) can now be 
related to $I_0$ by
\beq
\Delta^{\rm IR}(y,z) = \frac{\alpha}{\pi} \left[ I_0(y,z;M_\gamma)
+ \frac{1}{2} \Gamma(u,m^2_e,M_{\pi^+}^2;M_\gamma) \right] . 
\label{deltaIR}
\eeq

An analytic expression of the integral occurring in (\ref{decomp}) 
 was given in 
Appendix B of \cite{gin70} in terms of the quantities $U_i$:
\beq
\int\limits_0^{x_{\rm max}} dx \: \sum_{m,n} c_{m,n} I_{m,n} = 
\sum_{i=0}^7 U_i
 .
\eeq
As already noticed in \cite{CKNRT}, the quantity $J_9(i)$ given in 
Eq. (A9) of \cite{gin70} (which 
is needed for the evaluation of $U_7$) contains two  mistakes:
the plus-sign in the last line of (A9) should be replaced by a minus-sign, 
and $|\beta_i^{\rm max}|$ at the end of the first line of (A9) should 
simply read $\beta_i^{\rm max}$. 

The function $\Delta_1^{\rm IB}$ introduced in (\ref{deltas}) can now be 
obtained as
\beq \label{DeltaIB}
\Delta_1^{\rm IB} = \frac{2 \alpha}{\pi M_{\pi^+}^4} \sum_{i=0}^7 U_i 
\Big|_{\xi=0}.
\eeq
 
\section{Numerical analysis}
\label{sec: applic}
\renewcommand{\theequation}{\arabic{section}.\arabic{equation}}
\setcounter{equation}{0}

In the kinematically relevant range,
\beq
m_e^2 \le t \le (M_{\pi^+} - M_{\pi^0})^2 ,
\eeq
the $t$-dependence of the effective 
form factor can be approximated by the linear expansion
\beq \label{ex1}
f_{+}(t) = f_{+}(0) \ \left( 1 +  
 \frac{t}{M_{\pi^{\pm}}^2} \lambda_+ \right)  
\eeq
with an excellent degree of accuracy.
The observable decay rate
\beq 
\Gamma(\pi_{e3(\gamma)}) := 
\Gamma(\pi_{e3}) + \Gamma(\pi_{e3\gamma}) 
\eeq
can now be written as 
\beq
\Gamma(\pi_{e3(\gamma)}) = 
 {\cal N} \, \,  S_{\rm EW} (M_\rho,M_Z) \, 
\,  
|f_{+} (0)|^2  \, \, I(\lambda_+)  ,
\label{ex2}
\eeq
where 
\beqa
I (\lambda_+) &=&  \int\limits_{\cal D}  dy \, dz \, A_1 (y,z) 
\left( 1 + \frac{t}{M_{\pi^{\pm}}^2} \lambda_+ \right)^2 \nonumber \\
&=& a_0 + a_1 \, \lambda_+  + a_2 \,  \lambda_{+}^2  . 
\label{ex3}
\eeqa
In order to extract $|V_{ud}|$ we have to provide a theoretical
estimate of the form factor at $t=0$ and the phase-space integral.

\subsection{Numerical estimate of $\mbox{\boldmath $f_{+}(0)$}$}

In the isospin limit, $f_+(0)$ coincides with the vector form factor at 
zero momentum transfer, and is thus equal to 1, due to the 
conservation of the charged isospin current.
In the real world, all deviations from this value are 
therefore isospin suppressed.

At one-loop accuracy, the quantity $\wt{f}_+(0)$ obtained from 
the formula in (\ref{fplus-wiggle}) is unambiguously determined in terms 
of the masses of the pseudoscalar mesons. It deviates from 
1 by a tiny term quadratic in the isospin-breaking parameters 
$m_u - m_d$ and $e^2$ \cite{ademollo64}, 
\beq \label{fQCD}
\wt{f}_+(0) = 1 - 7 \times 10^{-6} .
\eeq
Further higher orders contributions to $\wt{f}_+(0)$ are 
negligibly small because of the aforementioned isospin suppression.

Therefore, the theoretical prediction of $f_+(0)$ requires a reliable 
estimate of the purely electromagnetic contribution $f_+|_{e^2 p^2}$.  
A numerical value for the coupling constant $K^r_{12}(\mu)$ entering in 
(\ref{fplus-em-loc}) has been given by Moussallam \cite{moussallam}: 
\beq K^r_{12} (M_\rho) = (-4.0 \pm 0.5)
\times 10^{-3} . \eeq 
For the (unknown) ``leptonic'' constants we resort
to the usual bounds suggested by dimensional analysis,
\beq |X_1|, \
|\wt{X}^r_6(M_\rho)| \leq 6.3 \times 10^{-3} .  
\eeq 
Using these numerical
values, (\ref{fplus-em-loc}) implies 
\beqa \label{fEM}
\left. f_{+} \right|_{e^2 p^2} &=& (4.6 \pm 0.1
\pm 0.4 \pm 0.3) \times 10^{-3} \nn 
&=& (4.6 \pm 0.5)  \times 10^{-3} .
\eeqa 
The three errors in the first line correspond to the uncertainties
of $K_{12}^r(M_\rho)$, $X_1$ and $\wt{X}_6^r(M_\rho)$, respectively. For
the final value, they have been added in quadrature. 

Despite the poor
present knowledge of the  ``leptonic'' constants, the uncertainty in 
the electromagnetic sector affects the final result for the effective form 
factor at zero momentum transfer by only $\pm 0.05 \%$,
\beq \label{thpred}
f_+(0) = 1.0046 \pm 0.0005 .
\eeq

\subsection{The phase space factor}

The theoretical prediction for the slope parameter 
\beq
\lambda_+ = M_{\pi^{\pm}}^2
\frac{d f_{+} (t)}{d t} \bigg|_{t=0}  . 
\eeq
is determined by the 
size of the low-energy constant $L^r_9$. With
\beq
L^r_9 (M_\rho) = (6.9 \pm 0.7) \times 10^{-3} 
\eeq
we find 
\beq 
\lambda_{+} = 0.037 \pm 0.003  
\eeq

\begin{table}
\renewcommand{\arraystretch}{1.2}
\caption{\label{table1} Coefficients entering the phase space integral}
\vspace{0.2cm}
\begin{center}
\begin{tabular}{|l|c|c|c|}
\hline
 & $a_0$ & $a_1$ & $a_2$ \\ \hline
$\alpha = 0$  & $7.375 \times 10^{-8}$ & $4.999 \times 10^{-11}$ & $
1.227 \times 10^{-14}$ 
\\ \hline 
$\alpha \neq 0$  & $7.383 \times 10^{-8}$ & $5.011 \times 10^{-11}$ & 
$1.230 \times 10^{-14}$ \\ 
\hline 
\end{tabular}
\end{center}
\end{table}

The numerical coefficients $a_{0,1,2}$ entering in the phase space 
expression (\ref{ex3})
are shown in Table \ref{table1}. Because of the smallness of the 
coefficients $a_{1,2}$, the final value of the Dalitz plot integral 
$I(\lambda_+)$ is practically insensitive to the exact size of the slope 
parameter and simply given by the parameter $a_0$.   
The inclusion of radiative corrections as described in Sec. \ref{sec: 
corr} increases the value of $I(\lambda_+)$ by only $0.1 \%$.

We have also evaluated numerically $a_0$ in the case corresponding to the 
fully inclusive one-photon decay (including the whole 4-particle phase 
space), finding no appreciable difference from the result in Table 
\ref{table1}.

\subsection{Determination of $\mbox{\boldmath $|V_{ud}|$}$}

The Kobayashi--Maskawa matrix element $|V_{ud}|$ can be extracted from the 
$\pi_{e3}$ decay parameters by
\beq \label{Vud}
|V_{ud}| = \frac{8 \, \, \pi^{3/2} \, \, 
\Gamma(\pi_{e3(\gamma)})^{1/2}}
{G_{\rm F} \, \, M_{\pi^+}^{5/2} \, \, S_{\rm EW}(M_\rho,M_Z)^{1/2} \, 
\, |f_+(0)| \, \, I(\lambda_+)^{1/2}} .
\eeq
We recall at this point that, according to our convention \cite{lept}, 
the Fermi coupling constant $G_F$ appearing in (\ref{Vud}) has to be 
identified with the muon decay constant. For the short distance 
enhancement factor we use the value \cite{MS93}
\beq
S_{\rm EW}(M_\rho,M_Z) = 1.0232,
\eeq
where leading logarithmic and QCD corrections have been included.
With the present $\pi^\pm$ mean life time \cite{PDG},
\beq
\tau_{\pi^\pm} = (2.6033 \pm 0.0005) \times 10^{-8} s,
\eeq
we finally obtain the relation 
\beq \label{Vud2}
|V_{ud}| = 9600.8 \, \left. \sqrt{ {\rm BR}(\pi_{e3(\gamma)}) } \right/ 
|f_+(0)| ,
\eeq
with an associated uncertainty
\beq \label{uncert}
\Delta |V_{ud}| = |V_{ud}| \left( \pm \frac{1}{2} \frac{\Delta
{\rm BR}}{\rm BR} \pm
\frac{\Delta f_{+}(0)}{f_{+}(0)} \right) .
\eeq

The present experimental precision for the branching ratio of the pionic 
beta decay cannot compete yet with the very small theoretical uncertainty 
in the determination of $V_{ud}$ generated by (\ref{thpred}). Using the 
latest value given by the Particle Data Group (PDG 2002) \cite{PDG},
\beq
{\rm BR} = (1.025 \pm 0.034) \times 10^{-8},
\eeq
together with (\ref{thpred}), we find
\beqa
|V_{ud}| &=& 0.9675 \pm 0.0160 ({\rm exp.}) \pm 0.0005 ({\rm theor.})\nn
         &=& 0.9675 \pm 0.0161 .
\eeqa 

However, a substantial improvement of the experimental accuracy is to be 
expected in the near future. 
The PIBETA experiment \cite{PSI} aims at measuring the branching 
ratio with a precision of about $0.5 \%$ in its current phase.
Inserting the present preliminary result obtained by the 
PIBETA Collaboration \cite{PSI},
\beq
{\rm BR} = (1.044 \pm 0.007 ({\rm stat.}) \pm 0.015 ({\rm syst.})) \times 
10^{-8},
\eeq
we find
\beqa
|V_{ud}| &=& 0.9765 \pm 0.0080 ({\rm exp.}) \pm 0.0005 ({\rm theor.})\nn
         &=& 0.9765 \pm 0.0080 ,
\eeqa
to be compared with the current PDG value \cite{PDG},
\beq
|V_{ud}| = 0.9735 \pm 0.0008 .
\eeq

\section{Conclusions}
\label{sec: Conclusions}
\renewcommand{\theequation}{\arabic{section}.\arabic{equation}}
\setcounter{equation}{0}

The present work was devoted to the study of the pionic
beta decay at the one loop level, with the order $\alpha$ 
radiative corrections to the amplitude included. We have been working
within the framework of the effective low-energy theory of
the standard model. Our main results in this respect 
are given by Eqs. (\ref{fplus-wiggle}) and (\ref{fplus-em-loc}).

We have discussed the influence of the various corrections on the 
determination of $\vert V_{ud}\vert$ from a high-precision 
measurement of the pionic beta decay rate. As far as strong
interaction corrections are concerned, the situation is most
advantageous, since the Ademollo-Gatto theorem requires 
the deviation of $f_+(0)$ from its value 1 in the isospin symmetric
limit to be quadratic in the meson mass differences 
$M_{\pi^+}^2-M_{\pi^0}^2$ and $M_{K^+}^2-M_{K^0}^2$.
This results in a very tiny correction at one loop, 
$\sim -7\times 10^{-6}$, and leads one to the expectation that higher
 order strong interaction corrections will not disturb this nice picture 
\footnote{The situation
here is very different from the case of the $K_{\ell 3}$ modes, where
at two loop one encounters ${\cal O}[(M_K^2-M_{\pi}^2)^2]$ 
counterterm contributions, which can have an influence on 
the determination of $\vert V_{us}\vert$, see e.g. the discussion in 
\cite{fks00}.}.

Electromagnetic corrections induced by the exchange of virtual
photons involve several unknown counterterms. However, naive
dimensional analysis indicates that their contribution also
remains small. For instance, they affect the extraction of 
$\vert V_{ud}\vert$ at the 0.05\% level only. However, they
represent the main source of theoretical error at present.
Putting together the short-distance corrections ($S_{\rm EW}$) and the 
long-distance corrections (to form factor and phase space), we estimate an 
overall radiative correction to the partial width of $ (+ 3.34 \pm 0.10) 
\% $ which is very close to other estimates \cite{jaus,Sirlin}. We stress 
that our number has been obtained within a completely model-independent 
framework for the long-distance corrections.

Thus, the pionic beta decay is very close to a theorist's
paradise, and a very precise prediction for its branching
ratio can be obtained. It remains to be seen whether the 
experimental progresses will eventually be able to reach
a comparable precision, and thus provide a very clean
and accurate determination of the Cabibbo angle.

\medskip

\noindent
{\small {\it Acknowledgements.} V. C. acknowledges partial support 
from MCYT, Spain, Grant No. FPA-2001-3031 and by ERDF funds from the 
European Commission. We are grateful to E. Frle\v{z} for valuable 
communications about the PIBETA experiment.} 

\section*{Appendix}

\appendix

\section{Photon Loop Functions}
\label{appA}
\renewcommand{\theequation}{\Alph{section}.\arabic{equation}}
\setcounter{equation}{0}

The photon loop contributions to the $\pi_{e 3}$ 
form factor depends on the electron mass $m_e$, the charged pion mass 
$M_{\pi^+}$ and the relativistic invariant 
$u = (p_+ - p_e)^2$. 
In order to express the loop functions in a compact way, 
it is useful to define the quantity
\begin{equation}     
 X = \frac{y - \sqrt{y^2 - 4 r_e}}{2 \sqrt{r_e}} , 
\end{equation} 
where $y$ has been defined in (\ref{defyz}).
In terms of $r_e, X, y$  and  the dilogarithm 
\begin{equation}
Li_2 (x) = - \int_{0}^{1} \frac{dt}{t} \log (1 - x t)   , 
\end{equation}
the functions contributing to $\Gamma (u,m_e^2,M_{\pi^+}^2 ; M_\gamma)$
are given by \cite{CKNRT}
\beqa \label{Gammac}
\lefteqn{\Gamma_c (u,m_e^2,M_{\pi^+}^2;M_\gamma)  =  
 2 M_{\pi^+}^2 y \, {\cal C} (u,m_e^2,M_{\pi^+}^2)} \nn *  
&& {} + 2  \log \frac{M_{\pi^+} m_e}{M_\gamma^2} 
\bigg(1 + \frac{X y 
\log X}{\sqrt{r_e} 
(1 - X^2)}  \bigg)  . 
\eeqa
\beqa
\lefteqn{{\cal C} (u,m_e^2,M^2)  =   \frac{1}{m_e M_{\pi^+}} \frac{X}{1 - 
X^2}} \nonumber \\*
&\times&
\left[  - \frac{1}{2} \log^2 X + 2 \log X \log (1 - X^2) - 
\frac{\pi^2}{6} + \frac{1}{8} \log^2 r_e \right. \nn
&&{}+ \left.  Li_2 (X^2) + Li_2  (1 - \frac{X}{\sqrt{r_e}}) + 
Li_2 (1 - X \sqrt{r_e})  \right]  ,  \nonumber \\*
&&
\eeqa 
\beqa
\Gamma_1(u,m_e^2,M_{\pi^+}^2) &=& \frac{1}{2} \Big[ -\,\ln r_e\,+\,
(4-3y){\cal F}(u,m_e^2,M_{\pi^+}^2) \Big]
\nonumber\\
\Gamma_2(u,m_e^2,M_{\pi^+}^2) &=& \frac{1}{2} 
\Big(1-\frac{m_e^2}{u}\Big) 
\Big[ - {\cal F}(u,m_e^2,M^2)(1-r_e)
\nonumber\\
&+& \ln r_e \Big]  -
\frac{1}{2}(3-y){\cal F}(u,m_e^2,M_{\pi^+}^2),
\eeqa
and
\beq
{\cal 
F}(u,m_e^2,M_{\pi^+}^2)\,=\,\frac{2}{\sqrt{r_e}}\,\frac{X}{1-X^2}\,\ln X.
\eeq

\section{Meson Loop Functions}
\label{appB}
\renewcommand{\theequation}{\Alph{section}.\arabic{equation}}
\setcounter{equation}{0}

The loop function $H_{PQ} (t)$ \cite{gl852,GL85} is given by 
\begin{equation} 
H_{PQ} (t) =\displaystyle\frac{1}{F_0^2} \bigg[ h_{PQ}^r (t,\mu) 
+ \frac{2}{3}t L_9^r(\mu) \bigg] ~, 
\end{equation} 
where
\begin{eqnarray} 
h_{PQ}^r (t,\mu) &=& 
\frac{1}{12 t} \lambda 
(t,M_P^2,M_Q^2) \, 
\bar{J}_{PQ} (t) \nn
&&{}+ \frac{1}{18 (4 \pi)^2} (t - 3 \Sigma_{PQ}) \nn
&&{}- \frac{1}{12} \bigg\{ \frac{2 \Sigma_{PQ} - t}{\Delta_{PQ}} [A_P(\mu) 
- A_Q(\mu)] \nn
&& \qquad {}- 2 [A_P(\mu) + A_Q(\mu)] \bigg\}  , 
\end{eqnarray} 
with
\beq
\lambda (x,y,z)  =   x^2 + y^2 + z^2  - 2 ( x y + x z + y z )  , 
\eeq
\beq 
\Sigma_{PQ}  =  M_P^2 + M_Q^2~, \qquad \Delta_{PQ}  =  M_P^2 -
M_Q^2   , 
\eeq
\beq
A_P(\mu)   =   - \frac{M_P^2}{(4 \pi)^2} 
\log \frac{M_P^2}{\mu^2}   , 
\eeq
and
\beqa
\lefteqn{\bar{J}_{PQ} (t)  = 
\frac{1}{32 \pi^2} \Bigg[ 2 + 
\frac{\Delta_{PQ}}{t} 
\log \frac{M_Q^2}{M_P^2} - \frac{\Sigma_{PQ}}{\Delta_{PQ}} 
\log \frac{M_Q^2}{M_P^2} }  \nn
&&{}  - \frac{\lambda^{1/2} (t,M_P^2,M_Q^2)}{t} \nn
&& \times
\log  \left( \frac{[t + \lambda^{1/2} (t,M_P^2,M_Q^2)]^2 - 
\Delta_{PQ}^2}{[t - 
\lambda^{1/2} (t,M_P^2,M_Q^2)]^2 - \Delta_{PQ}^2} \right) \Bigg]   .
\eeqa
The quantity $H_{PQ}(0)$ appearing in the evaluation of $f_+(0)$ is given 
by \cite{gl852}
\beqa \label{null}
H_{PQ}(0) &=& - \frac{1}{128 \pi^2 F_0^2} (M_P^2 + M_Q^2) \,
h_0 \! \left( \frac{M_P^2}{M_Q^2} \right), \nn
h_0(x) &=& 1 + \frac{2x}{1-x^2} \log x.
\eeqa
The chiral one-loop corrections comply with the Ademollo-Gatto theorem
through the property
\beq
H_{PQ}(0) =  - \frac{1}{192 \pi^2}\,\frac{\Delta_{PQ}^2}{F_0^2\Sigma_{PQ}}
\,+\,\cdots
\eeq
for $\Delta_{PQ}\ll\Sigma_{PQ}$.
For the theoretical determination of the slope parameter we need the 
derivative of the function $H_{PQ}(t)$ at $t= 0$ given by \cite{gl852}
\beqa \label{derH}
\left. \frac{d H_{PQ}(t)}{d t} \right|_{t = 0} &=&
\frac{2}{3 F_0^2} \left\{ L_9^r(\mu) - \frac{1}{128 \pi^2} \log \frac{M_P 
M_Q}{\mu^2} \right\} \nn
&&{} - \frac{1}{192 \pi^2 F_0^2} \, h_1 \! \left( 
\frac{M_P^2}{M_Q^2} \right) , \nn
h_1(x) &=& \frac{x^3 - 3 x^2 - 3 x + 1}{2 (x-1)^2} \log x \nn
&&{}
+\frac{1}{2} \left( \frac{x+1}{x-1} \right)^2 - \frac{1}{3}.
\eeqa

\end{document}